\documentclass[a4paper,11pt]{article}
\pdfoutput=1 

\usepackage{jheppub} 

\usepackage[T1]{fontenc} 

\usepackage{amsmath}
\usepackage{amssymb}
\usepackage{amsfonts}
\usepackage{breqn}
\usepackage[bottom]{footmisc}

\newcommand{\ISO}{\text{ISO}}

\newcommand{\SO}{\text{SO}}

\newcommand{\Diff}{\text{Diff}}
\newcommand{\dd}{\mathrm{d}}
\newcommand{\Lag}{\mathcal{L}}

\newcommand{\Reals}{\mathbb{R}}

\newcommand{\Mink}{\mathbb{M}}

\newcommand{\dS}{\text{d}\mathbb{S}}
\newcommand{\U}{\text{U}}

\newcommand{\equi}[1]{\left[#1 \right]_*}
\newcommand{\eqdef}{\stackrel{\scriptsize\textrm{def}}{=}}
\newcommand{\del}{\partial}

\title{\boldmath Inequivalent Goldstone Hierarchies for
Spontaneously Broken Spacetime
Symmetries}

\author[a]{Bernardo Finelli}

\affiliation[a]{Institute for Theoretical Physics,\\Utrecht University, Leuvenlaan 4, 3584 CE Utrecht, the Netherlands}

\emailAdd{b.finelli@uu.nl}

\abstract{The coset construction is a powerful tool for building theories that non-linearly realize symmetries. We show that when the symmetry group is not semisimple and includes spacetime symmetries, different parametrizations of the coset space can prefer different Goldstones as essential or inessential, due to the group's Levi decomposition. This leads to inequivalent physics. In particular, we study the theory of a scalar and vector Goldstones living in de Sitter spacetime and non-linearly realizing the Poincare group. Either Goldstone can be seen as inessential and removed in favor of the other, yet the thery can be healthy with both kept dynamical. The corresponding coset space is the same, up to reparametrization, as that of a Minkowski brane embedded in a Minkowski bulk, but the two theories are inequivalent.}

\begin{document} 
\maketitle
\flushbottom

\section{Introduction} \label{Introduction}
The concept of spontaneous symmetry breaking is a powerful organizational principle for effective theories found throughout different branches of physics. Because objects transforming under such symmetry group $G$ are not in a representation of $G$, the group is often said to be non-linearly realized. The coset construction -- so called because it relates the Goldstones bosons arising from the breaking pattern $G\to H$ to the coset space $G/H$ -- is part and parcel of building theories that non-linearly realize symmetry groups, the machinery for which was first introduced more than 50 years ago in \cite{coleman_structure_1969,callan_structure_1969}.

While the coset construction applied to internal (compact and semisimple) groups is well understood, considerable effort has been made in the last years to study its application to spacetime symmetry groups. After all, the breaking of these groups is at the heart of many concepts of physics, such as cosmology and condensed matter \cite{cheung_effective_2008,nicolis_zoology_2015,nicolis_spontaneous_2012,nicolis_relativistic_2014}. One of the most significant distinctions from internal groups is that a non-linear realization of spacetime symmetries can have fewer degrees of freedom than there are broken generators, the phenomenon of Goldstone inessentially \cite{volkov_phenomenological_1973,ivanov_inverse_1975,mcarthur_nonlinear_2010}.

Given spacetime coordinates $x$ and two kinds of Goldstones $\pi$ and $\xi$, we say $\pi$ is essential if \textit{(a)} it transforms without reference to $\xi$, that is, $\pi\to\tilde\pi(\pi,x)$; and \textit{(b)} $x$ and $\pi$ fully realize the symmetry, meaning their transformations depend on all group parameters. In this case, $\xi$ is unneeded to realize the group and can be discarded: it's inessential.

Goldstone inessentiality can also be formulated in terms of inverse Higgs constraints (IHCs). These are relations built out of the invariants that connect $\pi$ and $\xi$. If this relation includes derivatives of $\pi$ while $\xi$ appears only algebraically, the constraint can be solved to eliminate $\xi$ in terms of $\pi$ and its derivatives.\footnote{In some cases, different kinds of Goldstone might mix, yet an IHC still exists that allows one to be eliminated in favor of the other. The special galileon \cite{hinterbichler_hidden_2015} is an example. Nonetheless, a redefinition of coordinates and fields should unmix the Goldstones; see Section \ref{special galileon}.}

This hierarchy between Goldstones is peculiar, and the question then arises: is such structure unique? That is, given any parametrization (choice of coordinates) for some coset space, can we \textit{uniquely} determine the essentiality of a Goldstone boson? If true, it would mean the coset construction for spacetime symmetries is universal, meaning that different physicists arrive at the same theory regardless of how they choose to parametrize their coset spaces, assuming a common set of rules. On the other hand, the construction would be not unique if the two physicists end up eliminating different degrees of freedom from their theory and arrive at inequivalent actions.

We show the second case can happen. We provide a toy geometrical example in Section \ref{How to draw curves} and then a proper physical example in Section \ref{Poinc to Sitter}, where changing the parametrization changes the essential nature of the Goldstones, leading to different theories.

The reason is that the hierarchical property "$\pi$ transforms without reference to $\xi$" is a kind of structure. Reparametrizing the coset space induces a field redefinition between all objects that is guaranteed to preserve the group product structure, but not necessarily any other structure, including this hierarchical property.  Equivalently, each IHC will always be mapped to a new IHC, but this new constraint might be unusable if it can't be solved. This opens the door for physicists to make different choices of which Goldstones to eliminate (or which IHC to use). We discuss this in more detail in Section \ref{Coset construction universal?}.

In the literature, it's customary to perform the coset construction by means of the distinguished Maurer-Cartan form (e.g. \cite{low_spontaneously_2001,fels_moving_1998}). This is convenient, since we can refrain from explicitly deriving the transformation laws for our objects. But because those transformations are precisely our focus here -- in particular, which kinds of Goldstones transform without reference to the others -- we introduce in Section \ref{coset construction via norm} an alternative method based on \cite{fels_moving_1999} that requires computation of those transformations but dispenses use of the Maurer-Cartan form. Readers interested only in discussions and results are invited to skip it.

\paragraph{Conventions.} We work in $(-+++\ldots)$ signature. All transformations here are treated under the passive viewpoint, meaning we don't pullback the arguments of functions if the dependent variables transform, e.g., $x \to \tilde x(x)$ and $f(x) \to \tilde f(\tilde x)$. Unless said otherwise, quantities written in capital lettters (e.g., $X$ or $\Pi$) are invariant under the symmetry group under consideration.

\section{Non-uniqueness of coset construction}
First, we provide in Section \ref{How to draw curves} a toy geometrical example where the coset construction fails to deliver a unique result. For the interested reader, we discuss the more technical aspects of why this happens for non-semisimple groups in Section \ref{Coset construction universal?}.

\subsection{How to draw curves} \label{How to draw curves}

Suppose two physicists, Rachel and Leo, are asked to build the action for plane curves from the coset $\ISO(2)/\{1\}$, where $\{1\}$ is the trivial group containing only the identity. The two agree on the following common rules:

\begin{itemize}
    \item They will employ reparametrization invariance for their curves.
    \item The resulting action can contain only first or second derivatives.
    \item They should attempt to eliminate inessential degrees of freedom if possible.
\end{itemize}

\paragraph{Rachel's theory.} Rachel decides to parametrize the coset space as $\ell_R = e^{xP_1}e^{yP_2}e^{\theta J}$, where $P_i$ are translation generators and $J$ the rotation one. All the objects $\{x,y,\theta\}$ are functions of some diffeomorphism parameter $\lambda$. She computes the transformation laws of these objects under a Euclidean group element $(a^i,\varphi)$ and finds:

\begin{alignat}{4}
x &\to x \cos(\varphi) - y \sin(\varphi)+a^1, &&    &   \label{transformation of x}     \\ 
y &\to y \cos(\varphi) + x \sin(\varphi)+a^2, &&  & \label{transformation of y}\\
\theta &\to \theta + \varphi. &&  &
\end{alignat}

Rachel notices that just $x$ and $y$ are sufficient to fully realize the symmetry and that they transform without reference to $\theta$; she keeps them as essential and discards $\theta$ as inessential. Equivalently, she can find the constraint $\tan(\theta)\partial_\lambda x =\partial_\lambda y$, which is algebraic in $\theta$. She then computes the following action:

\begin{equation}
    S_R=\int\dd \lambda\sqrt{(x')^2+(y')^2}\,P\left(\frac{-y'x''+x'y''}{((x')^2+(y')^2)^{3/2}}\right),
\end{equation}

where $P$ is some arbitrary function and primes denote $\partial_\lambda$. This is of course familiar from plane geometry; the object inside the $P$ function is the extrinsic curvature $\kappa$ of a curve embedded in Euclidean space.

It is useful to understand how this action represents a prescription for drawing curves on paper. Rachel slides a ruler against the paper in a fixed direction. With the other hand, she holds a pen next to the ruler, allowing the pen to be pushed by it. Reparametrization invariance arises because the speed of the ruler can be removed as a degree of freedom; Rachel's actual degree of freedom is in moving the pen along the direction parallel to the ruler. For linear equations of motion obtained from $P(\kappa)=1$, she doesn't move the pen at all, only letting it be pushed by the ruler, and draws a straight line.

\paragraph{Leo's theory.} Leo then takes the parametrization Rachel used, but to be contrarian, flips the order of the exponentials, writing his as $\ell_L = e^{\theta J}e^{\sigma P_2}e^{\pi P_1}$. His transformation laws are:

\begin{alignat}{5}
\theta &\to \theta+\varphi, &&   &         \\
\pi &\to \pi +a^1\cos(\theta) + a^2 \sin(\theta), && \hspace{1.2cm} & \\ \label{transformation of pi and sigma}
\sigma &\to \sigma +a^2\cos(\theta) - a^1 \sin(\theta), && &
\end{alignat}

\pagebreak[3]

Leo now notices that $\theta$ and $\pi$ form an essential pair: together they fully realize the group, and neither one transforms with reference to the inessential $\sigma$. Leo then discards $\sigma$, which he could also do through the constraint $\sigma \partial_\lambda \theta = \partial_\lambda \pi$, and derives the following action:\footnote{Alternatively, he could've eliminated $\pi$ in favor of $\sigma$ but the final result is the same.}

\begin{equation}
S_L=\int\dd \lambda \,\theta'\,F\left(\pi+\frac{-\pi'\theta''+\theta'\pi''}{(\theta')^3}\right).
\end{equation}

\begin{figure}[t]
\centering
\includegraphics[scale=0.3]{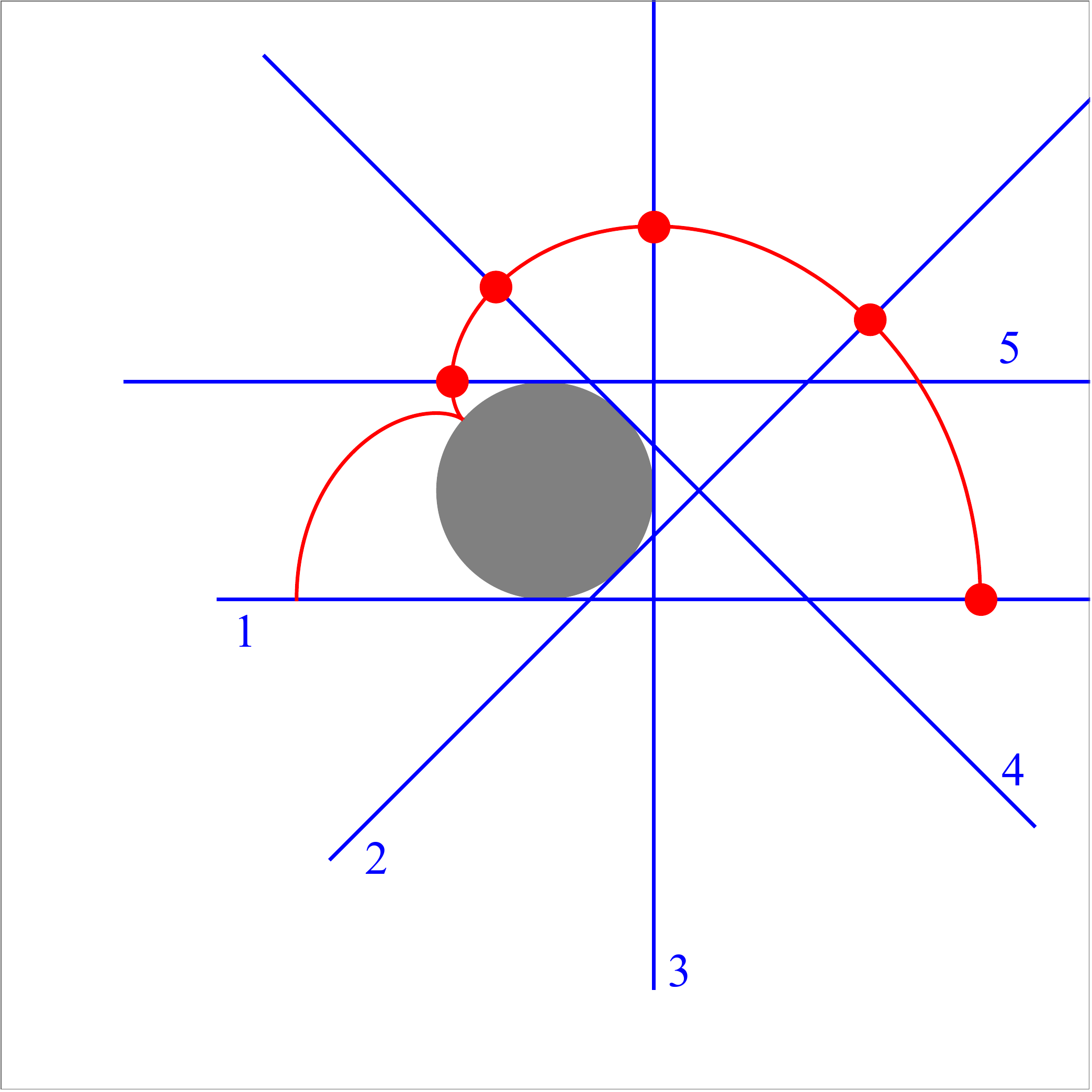}
\caption{To draw curves, Leo rolls a ruler without slipping around a fixed wheel while allowing the pen to be pushed by the ruler. In this picture, Leo has kept the pen still with respect to the ruler. The curve is drawn for a full rotation cycle, though we have included only five positions of the ruler to avoid overcrowding the figure.}
\label{fig:Rolling geometry}
\end{figure}

Leo's action is more peculiar. The object inside the $F$ function is a notion of torsion $\tau$,\footnote{Different from the torsion of spatial curves.} which we'll discuss shortly. In terms of drawing curves, it works as follows. He places a wheel together with a ruler on the paper. He then rotates the ruler \textit{without slipping} around the wheel, which is kept fixed. Again, the ruler's angular speed can be removed as a degree of freedom; the actual one is in moving the pen parallel to the ruler. For linear equations of motion obtained from $F(\tau)=\tau^2$, Leo doesn't move the pen with respect to the ruler and draws an involute of the wheel. See Figure \ref{fig:Rolling geometry} for visualization.\footnote{Readers familiar with children's toys will recognize this as a spirograph drawing.}

The invariant $\tau$ can be called torsion because it's connected to the winding of the pen around the wheel and thus to the displacement of the pen from its original position after one rotation cycle. A curve that doesn't close after one cycle must necessarily have nonzero torsion.

\paragraph{Inequivalence between the two.}

In Appendix \ref{plane geometry}, we show that no redefinition between Rachel's $(x,y)$ and Leo's $(\theta,\pi)$ exists. Even if we allow for higher derivatives in the action, the two can't be matched. They represent two distinct ways of drawing curves. While not particularly relevant for physics, the point of this example was to show that the coset construction doesn't necessarily produce unique results. Rachel and Leo started from the same coset space and employed the same prescription of removing inessential degrees of freedom, but arrived at inequivalent results.

In the following subsection, we discuss the reason why this can happen. In Section \ref{Poinc to Sitter} we apply the same logic Leo did to produce a physical example in spacetime.

\subsection{Why it's not unique} \label{Coset construction universal?}
We will first briefly review the connection of Goldstone bosons and homogeneous spaces in Section \ref{Goldstones = homo spaces} and the fundamentals of the coset construction in Section \ref{Homo spaces = coset spaces}. Then, in Section \ref{Additional structure}, we discuss inequivalencies that arise in non-semisimple groups.

\subsubsection{Goldstone bosons live in homogeneous spaces} \label{Goldstones = homo spaces}
Perhaps the fundamental property of Goldstone bosons is that a zero background value for one of these bosons can be transformed into a nonzero one under action of the symmetry group $G$:

\begin{equation}
    \pi =0 \quad \to \quad \pi \ne 0,
\end{equation}

so that the actual value of the Goldstone's vacuum is irrelevant. This property goes by the name of transitive group action. Isometry groups of maximally symmetric manifolds act transitively as well: the translations of Minkowski spacetime can move the origin somewhere else, and so can the homotheties (translations plus dilations) of de Sitter spacetime.\footnote{Recall the origin of de Sitter spacetime in conformal time is $-1/\text{Hubble}$, so not invariant under dilations, even though they act linearly.} The points of a maximally symmetry manifold are indistinguishable from each other.

The symmetries that move a field's background are said to be \textbf{broken}, while those that move the origin of spacetime are said to be \textbf{inhomogeneous}. If all \textit{all} inhomegeneous symmetries are \textit{unbroken}, we can collect all spacetime coordinates $x$ and all Goldstones $\pi$ into a single space $Q = \{x,\pi\}$ and the action of $G$ on $Q$ remains transitive. This $Q$ is then called a homogeneous space under $G$. Put another way, a homogeneous space has a single orbit under $G$: the whole space itself.

The $\pi$'s in this context should be seen as just coordinates of the space $Q$, not yet as functions of spacetime. A specific solution of the equations of motion is then the subspace given by the embedding $\pi(x)$.\footnote{As our focus is to build effective actions, we can always do so classically and then quantize the action afterwards. It would be interesting to extend the formalism discussed here to work with operators and Hilbert spaces from the get-go, though we won't do it here.}

\subsubsection{A homogeneous space is equivalent to a coset space \ldots} \label{Homo spaces = coset spaces}

The fact all coordinates and fields live in a homogeneous space turns out to be powerfully constraining for model building, because homogeneous spaces are mostly unique. To specify one such space under the group $G$, we simply need to know the \textbf{stability subgroup} $S$ of the origin of $Q$. That's the subgroup of $G$ that leaves \textit{both} the origin of spacetime \textit{and} the backgrounds of all fields invariant (i.e., unbroken homogeneous symmetries). Then the \textbf{orbit-stabilizer theorem} establishes that $Q$ corresponds to a coset space:

\begin{equation}
    Q \sim \frac{G}{S},
\end{equation}

where $\sim$ means equivalence in the sense that it preserves the group product (i.e., homeomorphism), but not necessarily any additional structure $Q$ might have. Concretely, this means that any $q\in Q$ can be written in terms of some group element $\ell s \in G$ acting on the origin of $Q$, where $\ell \in G$ is called a \textbf{lift} (or coset space representative) and $s\in S$ is an arbitrary stability element.

So for each element in $Q$ there's a corresponding element in $G$, with some $S$-ambiguity, thus intuitively $Q \sim G/S$. But a left coset space has a canonical group action by left multiplication of $g \in G$:

\begin{equation}
    \ell S \to \tilde \ell S \eqdef g \ell S.
\end{equation}

Hence, specifying the full symmetry group $G$ and the stability subgroup $S$ automatically specifies the transformation laws of all coordinates and fields, which by extension fixes all invariants that can be used to build an action. This construction -- mapping the physical entities in $Q$ to some coset space and deriving invariants -- goes by the name of \textbf{coset construction}.

\subsubsection{\ldots up to additional structure} \label{Additional structure}
The orbit-stabilizer theorem guarantees any homogeneous space $Q$ with a $G$-action is equivalent, up to additional structure, to the coset space $G/S$, the key phrase here being "up to additional structure."

Perhaps the first such structure one may think of is topology. This is a valid argument. Nonetheless, in the context of an effective theory, we typically are interested only in expanding fields perturbatively around the vacuum, so that the topology of field space is of little interest. Spacetime itself could have nontrivial topology as well, but by the same token we would prefer to restrict ourselves to local measurements that can't probe such exotica. So although topology might indeed lead to non-trivial physics beyond perturbation theory, we leave it aside in the following.

Typically in physics we have spacetime coordinates $x$ and internal space coordinates $\phi$. Crucially, spacetime is distinguished from internal space, because the $x$'s can only transform among themselves, i.e., $x\to\tilde x(x,g)$ under some $g\in G$. Another way of stating the same thing is that the isometries of a spacetime are intrinsic to the spacetime itself; they can't depend on what you put inside. On the other hand, the $\phi$ are allowed to mix with $x$, i.e., $\phi \to \tilde \phi(\phi,x,g)$; this is simply a non-uniform symmetry for our fields.

Similarly, if $Q$ contains essential Goldstones $\pi$ and inessential ones $\xi$, then by definition the action of $G$ only mixes the $\pi$ among themselves (and possibly the coordinates $x$) without reference to $\xi$.

This kind of hierarchy where some objects transform without reference to others is a type of additional structure.\footnote{It's a kind of fiber bundle structure, though somewhat different from how the concept is used in physics.} Not all field redefinitions will preserve it. For example, replacing $\pi$ by $\bar \pi = \bar \pi(\pi,\xi)$ will typically cause the transformation of $\bar \pi$ to also depend on $\xi$, and now the hierarchy is lost. Equivalently, a coset space parametrization in which an inverse Higgs constraint is algebraic can be mapped to one where the constraint is now differential and typically unsolvable.

This isn't particularly surprising. The orbit-stabilizer theorem only guarantees reparametrizations of the coset space preserve the group product structure, not this kind of hierarchical structure. But since this hierarchy is precisely linked to the removal of physical degrees of freedom, if multiple hierarchies exist then different physics can arise.

Here's one example where inequivalent hierachies for some coset space are possible. Following the Levi decomposition of $G$, any group can be written as $G = R \rtimes L$, so that each symmetry can be classified as belonging to either the radical $R$ or the (semi)simple factor $L$. See Appendix \ref{Levi decomposition} for details. Then when we parametrize the coset space, there are at least two orderings of the exponentials that endow the objects with different hierarchical structure; we leave the proof of this statement for Appendix \ref{Levi ordering}. The two Levi orderings are:

\begin{alignat}{3}
    \ell_R &\eqdef \text{radical symmetries}&&\times\text{simple symmetries},\\
    \ell_L &\eqdef \text{simple symmetries}&&\times\text{radical symmetries}.
\end{alignat}

In $\ell_R$, the radical objects fully realize the group, without reference to the simple ones. In $\ell_L$, the simple ones together with a reduced number of the radicals might fully realize the group. By means of example (both in Sections \ref{How to draw curves} and later in \ref{Poinc to Sitter}), we know that these orderings can swap which Goldstones are essential or inessential, so they can potentially lead to different physics for \textit{any} symmetry breakdown.

Of course, the above assumes $R$ even exists to begin with. If $G$ is a simple group, then the Levi decomposition is trivial and it's not clear if such ambiguities can arise. That they don't for internal groups is well established, but we don't know if that's the case for simple spacetime groups such as the conformal group $\SO(2,D)$ (see also \cite{klein_spontaneously_2017} for possible ambiguities in conformal group breakdown).

\section{Normalization construction} \label{coset construction via norm}

We now describe the general method for constructing objects that realize some symmetry, linearly or non-linearly. The procedure is essentially based on \cite{fels_moving_1998,fels_moving_1999}, and we direct the reader to those references for formal proofs of the method.

The reason why we use this technique rather than the usual one based on the Maurer-Cartan form is to highlight the importance of the transformation laws themselves, which dictate whether additional structure is present for the objects in our theory and, by extension, whether the resulting effective action is unique.

The basic idea is the notion that anything that can be transformed away by the symmetry group cannot, by definition, be an invariant. But the group is finite, so there's only a finite amount of quantities it can transform away before its symmetries have been used up. Anything that remains afterwards is an invariant.

For example, consider a nonrelativistic particle in 3D Euclidean space with position $\vec x$ under the Galilean group. We can spend all three translations moving $\vec x$ to $\vec 0$. Then we can spend all three boosts shifting the velocity $\dot {\vec x}$ to $\vec 0$. Now only rotations remain, but it's impossible to eliminate the acceleration $\ddot {\vec x}$ simply by rotating; at most, we can align it with some preferred axis. Thus $|\ddot {\vec x}|$ is what remains; it's the invariant of the Galilean group.

\paragraph{Step 1.} As input, one must inform the physicist about the full symmetry group of the problem, the broken symmetries, and the inhomogeneous symmetries in spacetime. This determines $G$ and its stability subgroup $S$.

\paragraph{Step 2.} Next, we must parametrize the lift of $G/S$. As discussed in Section \ref{Additional structure}, two convenient choices for non-semisimple groups are given by:

\begin{alignat}{3}
    \ell_R &\eqdef \text{radical symmetries}&&\times\text{simple symmetries},\\
    \ell_L &\eqdef \text{simple symmetries}&&\times\text{radical symmetries}.
\end{alignat}

\paragraph{Step 3.} We now derive the transformed lift $\tilde \ell$  under a group element $g\in G$, following the canonical group action on a coset via the group product: $\ell S\to\tilde\ell S= g\ell S$. This gives the transformation laws $x\to\tilde x$ and $\pi\to\tilde \pi$.

\paragraph{Step 4.} When deriving the transformation laws for the objects in the homogeneous space, we might observe that some transform without reference to the others, meaning the space might have some hierarchical structure. Suppose $Q=\{q,p\}$ with action under $g\in G$:

\begin{align}
    q &\to\tilde q (q;g), \\
    p &\to\tilde p(p,q;g),
\end{align}

where $\tilde q$ depends on \textit{all} parameters of $g$. Then we can construct a new homogeneous space $\bar Q\eqdef\{q\}$ deprived of the $p$'s, which still has a consistent action under $G$. Also, because $\tilde q$ depends on all parameters of $g$, this reduced space still realizes the full group $G$ (i.e., the action is faithful).

If this is possible then we can forget the $p$'s exist and perform the construction solely on the $q$'s, in which case the $p$'s are called inessential and the $q$'s essential. This is equivalent to imposing an inverse Higgs constraint after deriving the invariants, except here we do this from the very beginning.

Which objects are essential or inessential can depend on the ordering selected in step 2. This is because a reparametrization of the lift (i.e., a field redefinition) will not, in general, preserve the hierarchy between objects in the coset space.

\paragraph{Step 5.} The final step is to derive the actual invariants. At this point, we have a (possibly reduced in the previous step) homogeneous space with coordinates and fields, $Q=\{x,\pi\}$ and the transformation rules for $\tilde x$ and $\tilde \pi$ which follow from $\tilde \ell$. We now try to use the group action to set to zero as many objects in $Q$ as possible. 

Obviously $Q$ is an homogeneous space, so by definition everything in it can be eliminated. But we know how $x$ and $\pi$ transforms, so we know how $\partial_x \pi$ does too, as well as all higher derivatives. Thus we take our original homogeneous space $Q=\{x,\pi\}$ and extend it with a finite amount of derivatives $\partial_x \pi$, $\partial_x^2 \pi$, and so on.\footnote{Formally, the homogeneous space is a fiber bundle, so it can be prolonged into a jet bundle \cite[Ch.\ 4]{olver_equivalence_1995}.} We then transform those quantities under some special $g_* \in G$ to be determined later. These transformed objects are denoted with capital letters rather than tildes (e.g., $X$ instead of $\tilde x$) due to their special status as putative invariants.

Transforming $(x,\pi)$ under the $g_*$, we schematically have:

\begin{align}
    x   &                            \overset{g_*}{\to} X,      \\
    \pi &                           \overset{g_*}{\to} \Pi,   \\
    \frac{\del \pi}{\del x}  &   \overset{g_*}{\to} \frac{\dd \Pi}{\dd X}, \\
    \frac{\del^2 \pi}{\del x^2}  &   \overset{g_*}{\to} \frac{\dd^2 \Pi}{\dd X^2}, \\
     & \hspace{3mm} \vdots
\end{align}

\begin{enumerate}
\item We start by normalizing $X$ and $\Pi$ to zero,\footnote{We can set it to any constant without affecting the result. For clarity of notation, we set it to zero.} which allows us to solve for some of the parameters of $g_*$. If this completely fixes $g_*$, then ${\dd \Pi}/{\dd X}$ are the invariants of the theory,\footnote{Notice that setting $\Pi=0$ doesn't imply $\dd \Pi=0$.} and we're done.

\item If not, we then attempt to normalize as many of the ${\dd \Pi}/{\dd X}$ to zero as possible, which lets us fix more of the $g_*$. If $g_*$ is completely fixed by now, then the remaining ${\dd \Pi}/{\dd X}$ are the invariants we're after. If none remain, then the ${\dd^2 \Pi}/{\dd X^2}$ are the invariants.

\item If $g_*$ still hasn't been fixed, we repeat the procedure, setting as many of the ${\dd^2 \Pi}/{\dd X^2}$ to zero as possible, and so on. In the end, when $g_*$ is completely determined (which can always be done since the group is finite-dimensional, so a finite number of normalizations fixes all parameters), the lowest order in derivatives ${\dd^n \Pi}/{\dd X^n}$ that survived the process are the invariants.

\end{enumerate}
Invariant one-forms then follow by transforming the basis $\dd x$ under the $g_*$ found above. By extension we can build the invariant volume form $\dd V$:

\begin{align}
    \dd x   &                       \overset{g_*}{\to}    \dd X,      \\
    \dd V &\eqdef \frac{1}{D!} \dd X^0 \wedge \dd X^1 \wedge \ldots \wedge \dd X^D.
\end{align}

An invariant derivative can also be constructed, by inverting the invariant one-form as usual. That is, if $\dd X^\alpha = M^\alpha_{\,\beta}\dd x^\beta$, then the invariant derivative is $\dd/\dd X^\alpha = (M^{-1})^\beta_{\,\alpha}\partial_\beta$. Such derivatives
can act on the invariants we obtained to produce higher-order invariants, or act on additional matter fields that don't transform under the group $G$ and weren't part of the construction.

This gives the complete toolbox needed to build the most general invariant action.

Notice that in this procedure, we must use normalization constraints to fix all parameters of the group element $g_*$. However, in many cases, the objects we work with will transform linearly (i.e., in a representation) under some subgroup of $G$, typically the unbroken subgroup or the stability subgroup. In this case, if we contract objects covariantly under this subgroup, the corresponding group parameters will naturally drop out anyway. So these parameters don't need to be fixed, which saves us some time. But it's not always guaranteed that a certain lift parametrization will automatically induce a linear transformation that let us exploit covariance of objects. We encounter such issue in Section \ref{Poinc to Sitter}. 

\paragraph{Example: curvature of curves.}
Let's look at the quintessential example of planar curves. The coset space is $\ISO(2)/\{1\}$ and we parametrize the lift as $\ell_R = e^{xP_1}e^{yP_2}e^{\theta J}$. The three generators admit a matrix representation:

\begin{equation}
    P_1 = \begin{pmatrix} 0 & 0 & 1 \\ 0 & 0 & 0 \\ 0 & 0 & 0\end{pmatrix} \qquad
    P_2 = \begin{pmatrix} 0 & 0 & 0 \\ 0 & 0 & 1 \\ 0 & 0 & 0\end{pmatrix} \qquad
    J = \begin{pmatrix} 0 & -1 & 0 \\ 1 & 0 & 0 \\ 0 & 0 & 0\end{pmatrix},
\end{equation}

so that the group product can be easily computed in terms of matrix products. This gives the transformation laws that Rachel found in (\ref{transformation of x},~\ref{transformation of y}). Since $(x,y)$ fully realize the group, we discard $\theta$. We have the following quantities by transforming $x$, $y$ and derivatives of $y$ with respect to $x$, under the special group element $g_*=(a^i_*,\varphi_*)$:

\begin{align}
    X &= x \cos(\varphi_*)-y\sin(\varphi_*)+a_*^1,\\
    Y &= y \cos(\varphi_*)+x\sin(\varphi_*)+a_*^2,\\
    \frac{\dd Y}{\dd X} &= \frac{y' \cos(\varphi_*)+\sin(\varphi_*)}{ \cos(\varphi_*)-y'\sin(\varphi_*)}, \\
    \frac{\dd^2 Y}{\dd X^2} &= \frac{y''}{(\cos(\varphi_*)-y'\sin(\varphi_*))^3}.
\end{align}

Normalizing $X=Y=\dd Y/\dd X=0$ solves for the group element $g_*$:

\begin{equation}
    a^1_* = \frac{-x-yy'}{\sqrt{1+(y')^2}} \qquad a^2_* = \frac{-y + x y'}{\sqrt{1+(y')^2}} \qquad \varphi_* = -\arctan(y'),
\end{equation}

leaving us with the invariant curvature and measure:

\begin{equation}
    \frac{\dd^2 Y}{\dd X^2} = \frac{y''}{(1+(y')^2)^{3/2}} \qquad \dd X = \dd x\sqrt{1+(y')^2}.
\end{equation}

\subsection{Coordinate independence}
We can also perform the construction in a coordinate-independent manner. We declare all objects in the homogeneous space, both $x$ and $\pi$, to be functions of $D$ external diffeomorphism parameters $\lambda$. While $\lambda$ transforms under $\Diff(D)$, the basis forms $\dd \lambda$ transform under local $\text{GL}(D)$:

\begin{equation}
    \dd \lambda^a \to J^a_{\, b} \dd \lambda^b,
\end{equation}

where $J$ is the Jacobian of the diffeomorphism. Following a similar logic as before, we can transform $\dd \lambda$ under some special Jacobian $J_*$ to be determined to produce the invariant one-forms:

\begin{equation}
    \dd \lambda \overset{J_*}{\to} \dd \Lambda.
\end{equation}

Now, in addition to fixing the special group element $g_*$, we \textit{also} need to fix the special Jacobian $J_*$. That's $D^2$ extra parameters to fix! Luckily, because $\lambda$ is now our independent variable, we don't work with the quantities $\dd \Pi / \dd X$, but rather $\dd \Pi / \dd \Lambda$ and $\dd X / \dd \Lambda$:

\begin{align}
        \frac{\del x}{\del \lambda}  &  \overset{g_*,J_*}{\to}    \frac{\dd X}{\dd \Lambda}, \\
        \frac{\del \pi}{\del \lambda}  &  \overset{g_*,J_*}{\to}  \frac{\dd \Pi}{\dd \Lambda}.
\end{align}

The $\dd X/ \dd \Lambda$ now give precisely the extra quantities that can be normalized to convenient values in order to fix the Jacobian.

\paragraph{Example: diffeomorphic curvature of curves.}

As in the previous example, but now we impose diffeomorphism symmetry. Instead of taking derivatives with respect to the form $\dd x$, which gets transformed into $\dd X$ under the special group element $g_*$, we take derivatives with respect to the form $\dd \lambda$, which becomes $\dd \Lambda$ under the special Jacobian $J_*$. Importantly, $J$ is an element of \textit{local} $\text{GL}(1)$, so while $\dd g=0$, we have $\dd J \ne 0$. Thus, our quantities are:

\begin{align}
    X &= x \cos(\varphi_*)-y\sin(\varphi_*)+a_*^1,\\
    Y &= y \cos(\varphi_*)+x\sin(\varphi_*)+a_*^2,\\
    \frac{\dd X}{\dd \Lambda} &= \frac{1}{J_*}\left[x' \cos(\varphi_*)-y'\sin(\varphi_*)\right],\\
    \frac{\dd Y}{\dd \Lambda} &= \frac{1}{J_*} \left[y' \cos(\varphi_*)+x'\sin(\varphi_*)\right], \\
    \frac{\dd^2 X}{\dd \Lambda^2} &= \frac{J_*' \left(\sin (\varphi_* ) y'-\cos (\varphi_* ) x'\right)+J_* \left(\cos (\varphi_* ) x''-\sin (\varphi_* ) y''\right)}{J_*^3}, \\
    \frac{\dd^2 Y}{\dd \Lambda^2} &= \frac{J_* \left(\sin (\varphi_* ) x''+\cos (\varphi_* ) y''\right)-J_*' \left(\sin (\varphi_* ) x'+\cos (\varphi_* ) y'\right)}{J_*^3}.
\end{align}

Setting $X=Y=\dd Y/\dd \Lambda=0$ and $\dd X/\dd \Lambda = 1$ fixes everything:

\begin{equation}
    a_*^1 = \frac{-xx'-yy'}{\sqrt{(x')^2+(y')^2}} \qquad a_*^2=\frac{-yx'+xy'}{\sqrt{(x')^2+(y')^2}} \qquad \varphi_* = -\arctan(y'/x') \qquad J_* = \sqrt{(x')^2+(y')^2}, \label{diff curvature fix}
\end{equation}

so that the invariant curvature and measure are:

\begin{equation}
    \frac{\dd^2 Y}{\dd \Lambda^2} = \frac{-y'x''+x'y''}{[(x')^2+(y')^2]^{3/2}} \qquad \dd \Lambda = \dd\lambda \sqrt{(x')^2+(y')^2}.
\end{equation}

Note that $\dd^2 X/\dd \Lambda^2 = 0$ after imposing \eqref{diff curvature fix} so that the final number of invariant observables is the same as in the problem without diffeomorphism invariance. This is expected since coordinate independence is simply a redundancy in the description; the two problems are physically the same.

\subsection{Quasi-invariants}
The previous procedure concerns the construction of a strictly invariant action. Physics, however, isn't that strict and can tolerate actions that change by a total derivative. Terms that do so are called quasi-invariants, or Wess-Zumino terms.

To find these in $D$ spacetime dimensions, we must locate invariant $(D+1)$-forms $\beta$ that are exact, so $\beta=\dd \alpha$, but with $\alpha$ itself not being invariant. Then the invariance of $\beta$ together with $\dd^2=0$ imply the quasi-invariance of $\alpha$. And of course $\alpha$ is a $D$-form, so $\int \alpha$ will be a valid supplement to the action.

\pagebreak

We have invariant one-forms given by:

\begin{align}
    \dd x   &   \overset{g_*}{\to}     \dd X,      \\
    \dd \pi &   \overset{g_*}{\to}     \dd \Pi,   
\end{align}

evaluated under the special group element $g_*$ that we fixed before, and treating $\dd \pi$ as an independent form, that is, we don't write $\dd\pi^a = \partial_\mu\pi^a \dd x^\mu$. Higher forms can be constructed with sufficient applications of the wedge product between the $\dd X$ and $\dd \Pi$. The procedure is fairly standard, so we simply direct the reader to \cite{goon_galileons_2012} for more detailed instructions.

\section{Extended example: Poincare to de Sitter} \label{Poinc to Sitter}

Let's consider an extended example in spacetime and in higher dimensions that illustrates many of the ambiguities and inequivalencies that can arise when performing the coset construction for spacetime symmetry groups.

\paragraph{Step 1.} Suppose we are given the symmetry breaking pattern $\ISO(1,D) \to \SO(1,D)$. This covers the broken symmetries, but to fully determine the stability subgroup we need to know which symmetries are inhomogeneous in spacetime. There are two canonical options:

\begin{itemize}
    \item Since $\ISO(1,D)$ is the isometry group of Minkowski spacetime $\Mink^{D+1}$, we could take the spacetime origin to be the origin of $\Mink^{D+1}$. Then the inhomogeneous transformations are the translations, so that the overall stability group is $S=\SO(1,D)$.
    \item Since $\SO(1,D)$ is the isometry group of de Sitter spacetime $\dS^D$, we could take the spacetime origin to be the origin of $\dS^D$. Then the inhomogeneous transformations are the homotheties, so that the overall stability group is $S=\SO(1,D-1)$ (see the next step).
\end{itemize}

The choices are inequivalent; this is trivial to see since the first corresponds to the coset space $\ISO(1,D)/\SO(1,D)$ while the second to $\ISO(1,D)/\SO(1,D-1)$. The first gives rise to the usual embedding of the de Sitter hyperboloid in an ambient Minkowski space; as it has already been explored in \cite{goon_symmetries_2011}, we won't focus on it here. We will thus pick the second option, the coset space $\ISO(1,D)/\SO(1,D-1)$.

\paragraph{Step 2.} Let us now parametrize the lift that connects the elements in our theory to a coset in $\ISO(1,D)/\SO(1,D-1)$. Once again we are presented with inequivalent choices. One option would be to write:

\begin{equation}
    \ell_R = e^{x^\mu P^\Mink_\mu}e^{\pi P^\Mink_D}e^{\eta^\mu M^\Mink_{\mu D}}, \label{Lift for Mink brane in Mink bulk}
\end{equation}

where $P^\Mink$ are the usual translations of Minkowski space and $M^\Mink$ Lorentz transformations; the Greek indices $\mu,\nu$ range from 0 to $d=D-1$. But this lift parametrization gives the well known DBI action for a Minkowski brane embedded in Minkowski bulk \cite{goon_symmetries_2011}:

\begin{equation}
    S = \int \dd^D x\sqrt{1+(\partial \pi)^2}. \label{mink brane in mink bulk}
\end{equation}

Let's use the other Levi ordering, thus inverting the radical-then-simple order of \eqref{Lift for Mink brane in Mink bulk}. For clarity, we define a new basis for the simple generators:

\begin{align}
    D^{\dS} &= M^{\Mink}_{0D},\\
    P_i^{\dS} &= M^{\Mink}_{0i}-M^{\Mink}_{iD},\\
    M^{\dS}_{\mu\nu} &= M^{\Mink}_{\mu\nu},
\end{align}

with Latin indices $i,j$ ranging from 1 to $d$. The $D^{\dS}$ and $P_i^{\dS}$ generators satisfy the homothety algebra, that is, they are spacetime dilation and space translations, respectively:

\begin{equation}
    [P_i^{\dS},P_j^{\dS}] = 0 \hspace{8mm} [P_i^{\dS},D^{\dS}] = P_i^{\dS},
\end{equation}

so that the stability subgroup (unbroken group minus homotheties) is indeed $\SO(1,d)$, as advertised above. We thus write the simple-then-radical lift as:

\begin{equation}
    \ell_L = e^{x^i P^{\dS}_i} e^{t D^{\dS}}e^{\xi^\mu P^{\Mink}_\mu + \pi  P^{\Mink}_D}.
\end{equation}

\paragraph{Step 3.} We now need the transformation laws, which follow from the group action on a coset element: $\ell S \to  g \ell S$. While straightforward, the computation itself can be tedious (it helps to switch to conformal time $t = -\log(-\tau)$, with Hubble = 1). One concern we encounter is that then $\xi^\mu$ isn't a vector. For instance, it's strictly invariant rather than covariant under a dilation. In principle this isn't an issue; the procedure in Section \ref{coset construction via norm} doesn't require covariance under the unbroken subgroup. By inspection, though, we can see that a field redefinition $\xi^\mu = A^\mu /\tau$ gives the proper covariant transformation for $A^\mu$, so we will make use of this for simplicity. We stress, however, that this step is ad hoc; had we been unable to find this convenient field redefinition, we would have had to perform the full construction, without exploiting covariance.

Under the simple part of the group (unbroken $\SO(1,D)$) we obtain that $x^\mu=(\tau,x^i)$ transform as the (flat slicing) coordinates of de Sitter spacetime in conformal time (see Appendix \ref{dS isometries} for explicit expressions), $\pi$ as a scalar and $A^\mu$ as a vector.

\pagebreak[2]

As for the radical part of the group (broken $\Reals^{D+1}$), we have:

\begin{align}
    x^\mu &\to x^\mu, \\
    \pi &\to \pi - \theta, \label{pi transformation} \\
    A_\mu &\to A_\mu + \partial_\mu \theta,\label{A transformation} \\ 
    \theta &\eqdef \frac{1}{\tau}(c + b^i x^j \delta_{ij} + \frac{1}{2}a x^\mu x^\nu \eta_{\mu\nu}), \label{theta definition}
\end{align}

where $a,b^i,c$ are the parameters of the broken translations. Notice how $A^\mu$ transforms as if it were a gauge vector, with $\pi$ its longitudinal mode. However, we aren't interested in imposing gauge invariance, that is, for any choice of $\theta$, but rather only for the specific $\theta$ given above.

For convenience, it is useful to note that $\theta$ satisfies:

\begin{equation}
[\nabla_{(\mu} \nabla_{\nu)} + g_{\mu\nu}]\theta = 0,
\end{equation}

for $\nabla_\mu$ and $g_{\mu\nu}$ the usual geometrical objects of de Sitter space (in this context, $\theta$ is a scalar).

\paragraph{Step 4.} Our bosons are antisocial: $A_\mu$ transforms without $\pi$ and $\pi$ without $A_\mu$, and any by itself still fully realizes the broken translations (and the $x^\mu$ realize the rest of the group). This means we could, in principle, remove either one. For instance, removing the vector would give a dS galileon \cite{hinterbichler_non-linear_2012,goon_galileons_2012}. However, we're interested in investigating whether both can be kept (i.e., if no inverse Higgs constraint needs to be imposed, despite being available), so we will treat neither boson as removable.

\paragraph{Step 5.} To derive invariants, we first transform all objects under some special group element $g_*$, whose specific form will be fixed later:

\begin{align}
    x^\alpha &\overset{g_*}{\to}  X^\alpha,\\
    \pi &\overset{g_*}{\to}  \Pi,\\
    A_\beta &\overset{g_*}{\to} \mathcal A_\beta.
\end{align}

We wish to shift those objects back to the origin of spacetime and field space, so we set $X^0=-1$, $X^i=0$, $\Pi=0$ and $\mathcal A_\mu=0$. This solves for all the group parameters of the Minkowski translations $a,b^i$ and $c$, the dilation $\Lambda$, and the de Sitter translations $d^i$:

\begin{equation}
    a_*=A_0+\pi,\qquad \equi{b_i} = A_i,\qquad c_*=\frac{1}{2}(A_0-\pi) \qquad \Lambda_* = -\frac{1}{\tau} \qquad d_*^i = -x^i.
\end{equation}

The group parameters for the stability group $\SO(1,d)$ remain. However, both $\Pi$ and $\mathcal A_\alpha$, as well as derivatives $\dd/\dd X^\alpha$, transform covariantly under it, so we don't need to fix those parameters as long as we perform manifestly invariant contractions of the $\alpha,\beta$ indices.

Since we have exhausted the zeroth order objects $\Pi$ and $\mathcal A$, we extend it to their derivatives:

\begin{alignat}{3}
    \partial_\alpha \pi &\overset{g_*}{\to}   \frac{\dd \Pi}{\dd X^\alpha},\\
    \partial_\alpha A_\beta &\overset{g_*}{\to}   \frac{\dd \mathcal A_\beta}{\dd X^\alpha},
\end{alignat}

which must be evaluated under the $g_*$ we found above. 

\pagebreak

The covariant one-forms $\dd X$ are found in a similar manner, from transforming $\dd x$ under $g_*$. The result is:

\begin{align}
    \dd X^\alpha                                &= \frac{1}{\tau} \delta^\alpha_\mu \dd x^\mu,\\
    \frac{\dd \Pi}{\dd X^\alpha}                &= \tau \delta_\alpha^\mu (A_\mu + \partial_\mu \pi)\label{invariant scalar},\\
    \frac{\dd \mathcal A_\beta}{\dd X^\alpha}   &= \tau^2 \delta_\alpha^\mu\delta_\beta^\nu(\nabla_\mu A_\nu -\pi g_{\mu\nu}). \label{invariant vector}
\end{align}

These objects live in flat spacetime, so they must be contracted with $\eta_{\alpha\beta}$ or $\varepsilon_{\alpha_1\alpha_2\ldots}$. For ease of notation, we can write the corresponding objects living in the curved de Sitter space together with the volume measure, via the tetrad property $(\tau \delta^\mu_\alpha)(\tau  \delta^\nu_\beta) \eta^{\alpha \beta} = g^{\mu \nu}$:

\begin{align}
    \dd V &= \frac{1}{D!} \varepsilon_{\alpha_1\alpha_2\ldots}\dd X^{\alpha_1} \wedge \dd X^{\alpha_2} \wedge \ldots = \frac{\dd^D x}{\tau^D} = \dd^D x \sqrt{-g}, \\
    V_\mu &= A_\mu + \partial_\mu \pi, \\
    F_{\mu\nu} &= \partial_{[\mu} A_{\nu]} ,\\
    S_{\mu\nu} &= \nabla_{(\mu} A_{\nu)} - \pi g_{\mu\nu},
\end{align}

where $F_{\mu\nu}$ and $S_{\mu\nu}$ come from splitting \eqref{invariant vector} into its antisymmetric and symmetric parts, and $V_\mu$ is just \eqref{invariant scalar} renamed. The $\mu,\nu$ indices are to be contracted with $g^{\mu \nu}$. Note that $F_{\mu\nu}$ and $V_\mu$ are $\U(1)$-invariant, by accident as that wasn't part of the original construction, but $S_{\mu\nu}$ isn't.

There's another parametrization of these invariants that's more useful. By symmetrizing $\nabla_\mu V_\nu$, we can rewrite $S_{\mu\nu}$ without explicit reference to $A_\mu$:

\begin{equation}
    S_{\mu\nu} = \nabla_{(\mu}V_{\nu)} - [\nabla_{(\mu}\nabla_{\nu)} + g_{\mu\nu}]\pi.
\end{equation}

But $S_{\mu\nu}$ and $V_\mu$ are covariant so the following operator must be covariant as well:

\begin{equation}
    H_{\mu\nu} \eqdef [\nabla_{(\mu}\nabla_{\nu)} + g_{\mu\nu}]\pi.
\end{equation}

In principle, the strictly invariant action (no Wess-Zumino terms yet) then is:

\begin{equation}
    S = \int \frac{\dd^D x}{\tau^D} P(V_\mu,H_{\mu\nu};\nabla_\mu),
\end{equation}

where $F_{\mu\nu}$ is implicitly included given $V_\mu$ and $\nabla_\mu$. Making sure the action is healthy, however, further constrains it:

\begin{itemize}
    \item The scalar $\pi$ appears only in $H_{\mu\nu}$, with second derivatives. They need to appear in the special combinations that don't propagate ghosts \cite{goon_galileons_2012,bonifacio_shift_2019}, but we restrict the action to only first derivatives.
    \item Similarly, the vector $V_\mu$ can have kinetic terms of the form $\nabla_\mu V^\mu$ and $\nabla_{(\mu} V_{\nu)}$. Those, too, have to appear in a special combination that does not propagate a ghost, as a massive vector should have only three degrees of freedom. Such generalized Proca theories in curved spacetime have already been derived in \cite{tasinato_cosmic_2014,heisenberg_generalised_2017}; we must simply specialize to the case of de Sitter. This fixes the $V_\mu$ part of the action. 
\end{itemize}

After these considerations, the final strictly invariant Lagrangian is simply the generalized Proca one:

\begin{equation}
    \Lag_{\text{gen.Proca}}(V_\mu;\nabla_\mu),
\end{equation}

described in \cite{tasinato_cosmic_2014,heisenberg_generalised_2017}. Since the decomposition of the invariant $V_\mu$ in terms of the Goldstones is $V_\mu = A_\mu + \partial_\mu \pi$, this theory begs to be rewritten following the usual Stückelberg procedure. Defining the generalized Stückelberg Lagrangian, $\Lag_{\text{gen.Stück}}(A_\mu,\partial_\mu\pi) \eqdef \Lag_{\text{gen.Proca}}(A_\mu + \partial_\mu \pi)$, we write the action as:

\begin{equation}
    S=\int \dd^D x\sqrt{-g}\,\Lag_{\text{gen.Stück}}(A_\mu,\partial_\mu\pi;\nabla_\mu).
\end{equation}

It is not unusual that the strict invariants for $\pi$ ended up higher order in derivatives. From the transformation (\ref{pi transformation}), $\pi$ appears to be a galileon. As discussed in \cite{goon_galileons_2012}, galileon invariants tend to be higher order in derivatives. But now the theory has \textit{too much} symmetry. While it's technically $\ISO(1,D)$ invariant, that group gets drowned in the infinite $\U(1)$ gauge group. In order to rescue it while preserving the theory's health, we will use quasi-invariants that break $\U(1)$ but not $\ISO(1,D)$.

\subsection{Adding quasi-invariants} \label{WZ saved us}
In addition to strict invariants, we also have quasi-invariants, or Wess-Zumino terms, that change by a total derivative. For the sake of expediency, we just write down the first three, restoring the Hubble constant $H$:

\begin{align}
    W_1 &= \pi, \\
    W_2 &= (\partial \pi)^2 - DH^2\pi^2,\\
    W_3 &= (\square \pi) \left[ (\partial \pi)^2 -(D-1)H^2\pi^2\right]-\frac{2}{3}D(D-1)H^4\pi^3,
\end{align}

though we will require only $W_2$ for building a healthy theory. Notice these terms appear like the usual galileon operators, except with some corrections due to $H$. Indeed, in the limit $H \to 0$, we get the Minkowski galileons \cite{goon_galileons_2012,nicolis_galileon_2009} as expected from the group contraction.

Looking at $W_2$, the kinetic term has the wrong sign compared to the mass term. This is not a problem, because $(\partial \pi)^2$ also appears in the generalized Stückelberg Lagrangian, so we may hope to combine those two into something with the proper sign. Indeed, we can extract the mass term $-(m^2/2)(A_\mu+\partial_\mu\pi)^2$ from $\Lag_{\text{gen.Stück}}$ and add it to $q m^2 W_2$ where $q$ is some dimensionless constant. Performing the canonical normalization $\pi_c \eqdef m\sqrt{1-q}\pi$ then gives the following action:

\begin{dmath}
    S = \int \dd^D x \sqrt{-g} \left[-\frac{1}{4}F^2-\frac{1}{2}m^2 A^2 -\frac{m}{\sqrt{1-q}} (A \cdot \partial) \pi_c+ \Lag^{\text{int}}_{\text {gen.Stück}}\\ -\frac{1}{2}(\partial \pi_c)^2-\frac{1}{2}\left(\frac{q}{1-q}\right)DH^2 \pi_c^2 +\begin{matrix}\scriptsize\text{other}\\\scriptsize\text{WZ terms}\end{matrix}\right], \label{final action}
\end{dmath}

where $\Lag^{\text{int}}_{\text{gen.Stück}}$ denotes all generalized Stückelberg interactions. The theory is healthy as long as $0\le q<1$. Furthermore, the special case $q=0$ together with setting all other Wess-Zumino terms to zero restores the $\U(1)$ gauge symmetry.

To sum up, we have a healthy action \eqref{final action} constructed from the \textit{same} coset space as the action for a Minkowski brane embedded in Minkowski bulk \eqref{mink brane in mink bulk}, but the two theories have nothing to do with which other. In particular they don't even have the same number of degrees of freedom: here, the vector Goldstone can be kept, but in the Minkowski brane, it's inessential. Despite their coset constructions parametrizing the same coset space, they represent different physics due to flipping the Levi ordering of the parametrization.

\subsection{No inessential Goldstones} \label{No inessential GBs}
This problem is peculiar in that both Goldstones are essential and inessential: either $\pi$ or $A_\mu$ can be eliminated in favor of derivatives of the other. One way to see this is to return to the transformation laws (\ref{pi transformation}) and (\ref{A transformation}) and recall that the $\pi$ doesn't mix with the $A_\mu$ and vice-versa. We could've eliminated either and straightforwardly derived invariants using only one of them. It might also be instructive to look at this issue from the inverse Higgs constraint (IHC) point of view. Note that in this case the IHCs will \textit{not} be equivalent to integrating out fields via their equations of motion.

To eliminate $A_\mu$, we perform the following covariant normalization:

\begin{equation}
    V_\mu  = 0 \implies A_\mu = -\partial_\mu \pi.
\end{equation}

Alternatively, $\pi$ can be eliminated through the invariant normalization of $g^{\mu\nu}S_{\mu\nu}$:

\begin{equation}
    S  = 0 \implies \pi = \frac{1}{D}\nabla_\mu A^\mu. \label{IHC remove pi}
\end{equation}

For pure scalar theories with extended shift symmetries in de Sitter, see \cite{bonifacio_shift_2019}, while \cite{bonifacio_shift-symmetric_2019} discusses vectors with such symmetries in a similar context. Here we focused on showing that a theory mixing scalar and vectors is possible. Despite the existence of inverse Higgs constraints from a group-theoretical point of view, their usage isn't demanded by the physics.

\section{Discussion} \label{Discussion}
In this paper we have returned to the first principles of the coset construction to investigate its universality when applied to spacetime symmetry groups. We discussed the natural hierarchical structure that Goldstones can acquire when the group $G$ is not semisimple, which dictates which Goldstones are essential and which can be eliminated. By direct example, we showed that an arbitrary reparametrization of the coset space might not preserve this structure, changing which inessential Goldstones can be conveniently removed.

In particular, by reparametrizing the coset space of a Minkowski brane in Minkowski bulk, we constructed a theory for a scalar and vector Goldstones living in de Sitter space and non-linearly realizing the Poincare group. At first sight, one may wonder how this is possible, given that gauge Goldstones that non-linearly realize spacetime symmetries should not exist according to \cite{klein_no-go_2018} (our vector boson isn't gauge, but can of course be decomposed into one plus a scalar). Simply put, the assumptions of the no-go theorem aren't satisfied: our unbroken subgroup is de Sitter, whereas that of \cite{klein_no-go_2018} is Poincare, and \cite{klein_no-go_2018} assumes removal of the inessential Goldstones, which in our case can't be executed. Also, note that transformation for the vector is trivialized in the $H\to 0$ limit, so group contraction doesn't provide a counter-example to the no-go theorem. Indeed, the vector is an example of a symmetry realization unique to de Sitter without analogue in Minkowski, a possibility brought up in \cite{grall_symmetric_2019}.

That the universality of the coset construction isn't protected under transformations that change the Goldstone hierarchy isn't a new fact. In \cite{creminelli_inequivalence_2015}, the authors construct two inequivalent theories related by a map that mixes essential and inessential Goldstones, though they didn't bring up the hierarchy issue. It's not surprising these theories would then have different hierarchical structures.

A possible extension of this work would be to classify \textit{all} possible hierarchical structures for a given coset space with inequivalent physics. We provided two candidates, based on the Levi ordering, but we don't know if they are exhaustive. Furthermore, even if a reparametrization mixes essential and inessential Goldstones, it doesn't necessarily mean the resulting theories will be inequivalent; an example is given in \cite[Sec.\ 4.2]{klein_spontaneously_2017}.

Another avenue of further research is the study of simple spacetime symmetry groups. Nonlinear realizations of such groups can still involve inessential bosons. The basic example is how the breaking of the conformal group down to Poincare gives rise to an essential dilaton and an inessential special conformal boson \cite{hinterbichler_non-linear_2012}. Yet such groups have a trivial Levi decomposition, so wherever it is that their Goldstone hierarchy is coming from, it's not coming from there. Further study to determine whether the coset construction is unique in this case is required.

A last question concerns the issue of UV completion, which we haven't touched upon at all. Given two inequivalent theories derived from the same coset space, it would be interesting to see what the theories look like once the broken symmetries are restored and if they relate in any way.

\appendix
\appendix

\section{Rachel and Leo don't understand each other} \label{plane geometry}
Here we show that the two actions describing plane curves that Rachel and Leo found in Section \ref{How to draw curves} are inequivalent. 

Suppose they are equivalent and Rachel's language can be translated to Leo's and vice-versa. Then, we have to find an invertible redefinition that maps Rachel's variables to Leo's. A naive first attempt would be to set $\ell_R = \ell_L$, then use the inverse Higgs constraints that Rachel and Leo found to eliminate the inessential fields in terms of derivatives of the essentials. However, this redefinition would mix fields with their own derivatives, and one straightforwardly shows they are not invertible. For example, mapping Rachel to Leo then back to Rachel doesn't output the original input.

Instead, notice that Rachel's $y$ transforms as a shift under $P_1$, that is, $y \to y+a^1$, but Leo's corresponding $\pi$ doesn't:

\begin{equation}
    \pi \to \pi+ a^1 \cos\theta,
\end{equation}

while $\theta$ doesn't transform. To try to match the two actions, we must redefine $\pi = \pi(\phi,\theta)$ so that $\phi$ transforms as a shift under $P_1$ as well. Notice the definition of $\theta$ is irrelevant due to the gauge symmetry. To find the $\pi$ redefinition, we consider the infinitesimal variation $\delta  \pi(\phi,\theta)$ together with the conditions $\delta\phi = a^1$ and $\delta \theta = 0$ which gives the differential equation:

\begin{equation}
    \partial_\phi \pi = \cos\theta,
\end{equation}

solved by $\pi = \phi \cos\theta + f(\theta)$ for $f$ a free function. If we now take Leo's fundamental invariant torsion and replace $\pi$ by $\phi$, we get:

\begin{align}
    \tau &= \pi+\frac{-\pi'\theta''+\theta'\pi''}{(\theta')^3},\\
         &= \frac{\cos\theta}{(\theta')^2}\phi''-\frac{2(\theta')^2\sin\theta  +\theta''\cos\theta }{(\theta')^3}\phi' +\left(f''(\theta)+f(\theta)\right)
\end{align}

Now, we need to pick a gauge for $\theta(\lambda)$ together with the function $f$ so that the above becomes the same mathematical expression for Rachel's curvature invariant,

\begin{equation}
    \kappa =  \frac{y''}{[1+(y')^2]^{3/2}}, \label{Rachel with gauge choice}
\end{equation}

where we picked the $x(\lambda)=\lambda$ gauge for Rachel to simplify the expression. At the same time, we have to pick a gauge for $\theta(\lambda)$ so that Rachel's and Leo's invariant line elements also match:

\begin{align}
\dd \ell_R &= \sqrt{1+(y ')^2} \dd\lambda \\
\dd \ell_L &= \theta' \dd\lambda 
\end{align}

This is impossible; matching the line elements requires $\theta = \int \dd\lambda \sqrt{1+(y')^2}$ which causes Leo's torsion to become nonlocal while Rachel's is local. Notice we never invoked any action, thus the argument holds even if Rachel and Leo are allowed to use higher derivative invariants.

\section{Levi decomposition}\label{Levi decomposition}
Let us quickly recall the notion of the Levi decomposition. Any finite Lie group $G$ can be decomposed\footnote{To make the claim mathematically unimpeachable, we note the decomposition also admits discrete group factors and that the simple factor is only unique up to conjugation by the group's nilradical. We ignore these technicalities here.} using a single semidirect product \cite{oraifeartaigh_lorentz_1965}: 

\begin{equation}
    G = R \rtimes L,
\end{equation}

where:

\begin{itemize}
    \item $L$ is the \textbf{simple (or Levi) factor}, a semisimple group;
    \item $R$ is the \textbf{radical}, a group whose algebra is maximally solvable.
\end{itemize}

A subalgebra $\mathfrak i$ of $\mathfrak g$ is solvable if it's an ideal (so $[\mathfrak i,\mathfrak g]\subseteq\mathfrak i$) and if it telescopes to zero upon repeated application of the commutator, meaning:

\begin{alignat}{3}
    [\mathfrak i,\mathfrak i] &= \mathfrak i_1 && \subset \mathfrak i, \\
    [\mathfrak i_1,\mathfrak i_1] &= \mathfrak i_2 && \subset \mathfrak i_1, \\
    [\mathfrak i_2,\mathfrak i_2] &= \mathfrak i_3 && \subset \mathfrak i_2, \\
    & \vdots && \\
    [\mathfrak i_n,\mathfrak i_n] &= 0,
\end{alignat}

after a finite number $n$ of steps. The largest such ideal is then the algebra's radical. 

Some examples of Levi decomposition:

\begin{itemize}
    \item Poincare $\sim$ translations $\rtimes$ Lorentz,
    \item Galileo $\sim$ (translations and boosts) $\rtimes$ rotations,
    \item General affine $\sim$ (translations and dilation) $\rtimes$ special linear.
\end{itemize}

\subsection{Levi ordering} \label{Levi ordering}
Now suppose we want to work with the coset space $G/S$. We will assume that $G$ is not semisimple, but $S$ is. Ideally we would like to classify all possible hierarchical structures the homogeneous space could have, but we will limit ourselves to only showing at least two inequivalent ones exist.

Let $P$ be the generators living in the radical of $G$, and $T$ those in the simple factor. Finally, if a simple generator is not in the stability group $S$, denote it by $A$; conversely, denote a simple generator in $S$ by $V$. Then at least two parametrizations of the lift give different structures:

\begin{alignat}{3}
    \ell_R &\eqdef e^{z P}e^{\xi A} &&\hspace{5mm}(z\text{ transforms by itself)},\\
    \ell_L &\eqdef e^{\xi A}e^{z P} &&\hspace{5mm}(\xi\text{ transforms by itself)}.
\end{alignat}

To establish this, we act with the group $G$ and check the form of each transformation law. We will make use of the following braiding identities:

\begin{align}
    e^{aP}e^{uT} &= e^{uT}e^{M_uaP},\\
    e^{aP}e^{vT} &= e^{vT}e^{M_vaP},\\
    e^{aP}e^{\xi A} &= e^{\xi A}e^{v(\xi)V}e^{M_\xi aP},
\end{align}

where $M_u$ is a linear map. They follow from the Baker–Campbell–Hausdorff formula upon usage of the ideal property of the radical, $[P,T]\sim P$, together with closure of the $T$'s and $V$'s.

\paragraph{Radical first.} We act with some $g=e^{uT}e^{aP}$ on $\ell_R S$ to derive transformation laws:

\begin{align}
    \tilde \ell_R S &= e^{uT}\underline{e^{aP}e^{zP}}e^{\xi A}S \\
                    &= \underline{e^{uT}e^{\tilde z(z;a)P}}e^{\xi A}S \\
                    &= e^{M_u\tilde z(z;a)P}\underline{e^{uT}e^{\xi A}}S \\
                    &= e^{M_u\tilde z(z;a)P}e^{\tilde \xi(\xi;u)A}S,
\end{align}

where we used, in order, closure of $R$, then braiding, then closure of $L$; the underline denotes where we use each argument. To conclude, the objects in the homogeneous space transform as:

\begin{equation}
    z \to  M_u \tilde z(z;a), \hspace{5mm} \xi \to \tilde \xi(\xi;u),
\end{equation}

so that $z$ fully realizes the group: it transforms by itself and its transformation depends on all group parameters.

\paragraph{Simple first.} Without loss of generality, flip the order of the group element, so now $g=e^{aP}e^{uT}$ acting on $\ell_L S$:

\begin{align}
    \tilde \ell_L S &= e^{aP}\underline{e^{uT}e^{\xi A}}e^{z P}S \\
                    &= e^{aP}e^{\tilde \xi(\xi;u)A}\underline{e^{\tilde u(u;\xi)V}e^{zP}}S \\
                    &= \underline{e^{aP}e^{\tilde \xi(\xi;u)A}}e^{M_{\tilde u}z P}S \\
                    &= e^{\tilde \xi(\xi;u)A}\underline{e^{v(\xi;u)V}e^{M_{\tilde \xi}aP}e^{M_{\tilde u}zP}}S \\
                    &= e^{\tilde \xi(\xi;u)A}e^{\tilde z(z,\xi;u,a)P}S,
\end{align}

using closure of $L$, then braiding, then braiding again, then closure of $P$ with $V$. Thus the transformations are:

\begin{equation}
    \xi \to \tilde \xi(\xi;u),  \hspace{5mm} z\to\tilde z(z,\xi;u,a), 
\end{equation}

so $\xi$ transforms by itself, but $z$ transforms with reference to $\xi$.

\subsection{Example: Special galileon} \label{special galileon}
Let us use the techniques of the Levi decomposition and ordering to study the hierarchy of a particular kind of theory, the special galileon \cite{hinterbichler_hidden_2015}. The usual galileon theory is built out of the coset space $\text{GAL}(D,1)/\SO(1,D-1)$ parametrized by:

\begin{equation}
    \ell = e^{x^\mu P_\mu}e^{\phi Q}e^{\xi^\mu B_\mu},
\end{equation}

with the only nonzero commutator (besides those with Lorentz $M_{\mu\nu}$, which follow from a generator's tensorial structure) being:

\begin{equation}
    [P_\mu,B_\nu] = \eta_{\mu\nu}Q.
\end{equation}

One then derives that $\xi_\mu$ is inessential, in the sense we defined before: the transformations of $x^\mu$ and $\phi$ don't depend on $\xi^\mu$ while fully realizing the symmetry group.

The algebra above can be extended with an additional generator $S_{\mu\nu}$, symmetric in indices, such that \cite{hinterbichler_hidden_2015,garcia-saenz_gauged_2019}:

\begin{align}
    [P_\mu,S_{\nu\rho}] &= \eta_{\mu\nu}B_\rho +  \eta_{\mu\rho}B_\nu -\frac{2}{D}\eta_{\mu\rho}B_\mu, \\ \label{sgal alegbra}
    [B_\mu,S_{\nu\rho}] &= \alpha^2 \left(\eta_{\mu\nu}P_\rho +  \eta_{\mu\rho}P_\nu -\frac{2}{D}\eta_{\mu\rho}P_\mu\right),\\
    [S_{\mu\nu},S_{\rho\sigma}] &= \alpha^2 \left( \eta_{\mu\rho}M_{\nu\sigma} + \eta_{\mu\sigma}M_{\nu\rho}+\eta_{\nu\rho}M_{\mu\sigma}+\eta_{\nu\sigma}M_{\mu\rho} \right).
\end{align}

One then performs the coset construction paremetrizing the coset space lift as:

\begin{equation}
        \ell = e^{x^\mu P_\mu}e^{\phi Q}e^{\xi^\mu B_\mu}e^{\frac{1}{2}\sigma^{\mu\nu}S_{\mu\nu}},
\end{equation}

for which now one obtains that the $S_{\mu\nu}$ mixes $x^\mu$, $\xi^\mu$ and $\phi$ together. It would appear the addition of a new symmetry has rendered $\xi_\mu$ essential. Nonetheless, such vector can still be removed by means of inverse Higgs constraint \cite{garcia-saenz_gauged_2019}. We will show, however, that a different parametrization exists, one in which the Goldstone hierarchy is manifest and there is no mixing with $\xi^\mu$.

To do this, we must Levi decompose this new algebra, using the methodology outlined in this appendix. First, recall the standard Galileo group, without the $S_{\mu\nu}$, has the decomposition:

\begin{equation}
    \text{GAL}(D,1) = \exp(\text{span}(P,Q,B)) \rtimes \SO(1,D-1).
\end{equation}.

But $\text{span}(P,Q,B)$ is still an ideal even when adding $S_{\mu\nu}$:

\begin{equation}
    [\text{span}(P,Q,B),S] = \text{span}(P,Q,B).
\end{equation}

So it is still contained within the radical of the special Galileo group; the addition of the $S_{\mu\nu}$ doesn't reduce the original radical in size. 

It doesn't increase it either. This is easy to see, since $[S,S]\sim M$ violates the solvability criterion. Hence, the new generators must be added to the simple factor, the Lorentz group. And due to nontrivial commutators between $S_{\mu\nu}$ and $M_{\mu\nu}$, it must genuinely enlarge the Lorentz group, rather than just being an additional direct factor.

Simple algebras are well classified, and one can immediately check then that $\text{span}(M,S)$ is the special linear algebra. A way to intuit this is that the parameters of a Lorentz transformation can be collected in an antisymmetric tensor $a_{\mu\nu}$, those of the additional symmetries in a traceless symmetric tensor $s_{\mu\nu}$ \cite{hinterbichler_hidden_2015}; the two together give any traceless matrix. Hence, the special Galileo group decomposes as:

\begin{equation}
\text{SGAL}(1,D) = \exp(\text{span}(P,Q,B)) \rtimes \text{SL}(D).
\end{equation}

There must exist a choice of basis for which the radical transforms under the simple factor as a representation. The special Galileo algebra as given in the basis of \ref{sgal alegbra} does not display this property, so let us perform a basis change:

\begin{equation}
    \tilde P_\mu = \frac{1}{\sqrt{2}}\left({\alpha}P_\mu + B_\mu\right), \qquad \tilde B_\mu = \frac{1}{\sqrt{2}}\left(B_\mu -{\alpha}P_\mu\right), \qquad \tilde Q = \alpha Q, \qquad \tilde S_{\mu\nu} = S_{\mu\nu}/\alpha.
\end{equation}

The special Galileo algebra now reads:

\begin{align}
    [\tilde P_\mu,\tilde B_\nu] &= \eta_{\mu\nu}\tilde Q,\\
    [\tilde P_\mu,\tilde S_{\nu\rho}] &= \eta_{\mu\nu}\tilde P_\rho +  \eta_{\mu\rho}\tilde P_\nu -\frac{2}{D}\eta_{\mu\rho}\tilde P_\mu \\
    [\tilde B_\mu,\tilde S_{\nu\rho}] &= -\eta_{\mu\nu}\tilde B_\rho -  \eta_{\mu\rho}\tilde B_\nu +\frac{2}{D}\eta_{\mu\rho}\tilde B_\mu,\\
    [\tilde S_{\mu\nu},\tilde S_{\rho\sigma}] &=  \eta_{\mu\rho}M_{\nu\sigma} + \eta_{\mu\sigma}M_{\nu\rho}+\eta_{\nu\rho}M_{\mu\sigma}+\eta_{\nu\sigma}M_{\mu\rho} .
\end{align}

Now, if we parametrize the coset space as

\begin{equation}
    \ell = e^{x^\mu\tilde P_\mu}e^{\phi\tilde Q}e^{\xi^\mu \tilde B_\mu}e^{\frac{1}{2}\sigma^{\mu\nu}\tilde S_{\mu\nu}},
\end{equation}

the arguments of Section $\ref{Levi ordering}$ guarantee that $x^\mu$, $\xi^\mu$ and $\phi$ transform as representations under $\tilde  S_{\mu\nu}$. As usual, the condition $[\tilde S,\tilde P]\sim \tilde  P$ implies the braiding,

\begin{equation}
e^{s \tilde  S}e^{x \tilde  P} = e^{M_s x \tilde  P}e^{s\tilde  S},
\end{equation}

for some linear map $M_s$, and a similar argument for $\xi$ and $\phi$. Thus, they don't mix under action of $S_{\mu\nu}$. In particular, under the entire group, $x$ transforms only with itself, and $\phi$ transforms with $\phi$ and $x$, and together they fully realize the group. The $\phi$ is then an essential Goldstone, in the group action definition we used through this work.

\subsection{Supergroups}
While we have not discussed supersymmetries in this work, let us briefly outline how the Levi decomposition applies to them, as this is relevant for understanding the hierarchy of Goldstone \textit{fermions} arising in broken supersymmetric theories.

First, recall that a super Lie group is still a group \cite{yagi_super_1993}. It is still a manifold with morphisms satisfying certain axioms, but the manifold is allowed to have both bosonic coordinates $x^\mu$ and fermionic coordinates $\theta^\alpha,\bar \theta^{\dot \beta}$. Much in the same way $P_\mu$ generates translations of $x^\mu$, the supercharges $Q_\alpha, \bar Q_{\dot \beta}$ generate supertranslations of $\theta^\alpha, \bar\theta^{\dot \beta}$ \cite{laszlo_unification_2017}.

A super Lie group, being a group, must satisfy the Levi decomposition theorem, and thus one needs to ask if the supertranslations go in the radical or the simple factor. For the super Poincare group, one can realize the algebra as a standard algebra of commutators, rather than a graded one \cite{ferrara_supergauge_1974,salam_supergauge_1974}, then check for the solvable and ideal properties of the radical. Perhaps unsurprisingly, the supertranslations go in the radical \cite{laszlo_unification_2017}:

\begin{equation}
    \text{sISO}(1,D) = \exp(\text{span}(P,Q,\bar Q)) \rtimes \SO(1,D-1).
\end{equation}

Much of the machinery described in this appendix referring to coset space parametrization and hierarchy of Goldstone fields applies unchanged when supergroups are seen in this manner.\footnote{Under this view, the super Poincare group dodges the Coleman-Mandula theorem because its radical is enhanced, which is one of the few permitted ways to nontrivially extend the Poincare group as per O'Raifeartaigh's theorem. \cite{oraifeartaigh_lorentz_1965}}

\section{Finite de Sitter isometries} \label{dS isometries}
The transformation laws for $x^\mu=(\tau,x^i)$ derived in Section \ref{Poinc to Sitter} correspond to the de Sitter isometries. The straightforward ones are the translations $d^i$, dilation $\Lambda$ and rotations $\theta_{ij}$:

\begin{align}
    \tau &\to \Lambda \tau, \\
    x^i &\to \Lambda R(\theta)^i_{\,j}(x^j+d^j).
\end{align}

Meanwhile, a boost with rapidity $\beta$ along the $x$-direction is given by:

\begin{align}
    \tau &\to \frac{2 \tau }{(1-H^2 \eta_{\mu\nu}x^\mu x^\nu) +(1+H^2 \eta_{\mu\nu}x^\mu x^\nu) \cosh(\beta)+2 H x \sinh (\beta )}, \\
    x &\to \frac{2 x \cosh(\beta)+\sinh(\beta)(1+H^2 \eta_{\mu\nu}x^\mu x^\nu)/H }{(1-H^2 \eta_{\mu\nu}x^\mu x^\nu) +(1+H^2 \eta_{\mu\nu}x^\mu x^\nu) \cosh(\beta)+2 H x \sinh (\beta )}, \\
    y^j &\to \frac{2 y^j }{(1-H^2 \eta_{\mu\nu}x^\mu x^\nu) +(1+H^2 \eta_{\mu\nu}x^\mu x^\nu) \cosh(\beta)+2 H x \sinh (\beta )},
\end{align}

for $j\ne 1$. Boosts along the other $y^j$-directions follow identically by rotational symmetry. Upon changing to physical time and taking the $H\to0$ limit, we retrieve the Minkowski boost.

\acknowledgments

The author would like to thank Sadra Jazayeri, Tanguy Grall,  Diederik Roest and David Stefanyszyn for useful discussions, Guilherme Pimentel for suggestion of future work, and particularly Garrett Goon and Enrico Pajer for commentary on the draft of this work. BF is supported by the the Netherlands Organisation for Scientific Research (NWO) that is funded by the Dutch Ministry of Education, Culture and Science (OCW).


\bibliographystyle{ieeetr}
\bibliography{references.bib}



\end{document}